# Signs of Lorentz violation in electrodynamics: variable speed of light and the photon mass

ROBERTO ASSUMPÇÃO

*PUC-Minas, Av. Pe. Francis C. Cox, Poços de Caldas,*
*MG  37701-355, Brasil*
assump@fem.unicamp.br

Recent studies of spacetime anisotropy in the context of local Lorentz invariance (LLI) based on classical Michelson-Morley experiments, as well Kennedy-Thorndyke tests, pointed out the existence of  terms first order in *v*/c and of angular signatures independent of *v*. This contribution replaces the Lorentz symmetry by a velocity gauge transformation following an argument centred on observability. Results show even and odd order terms and indicate that motion is always underestimated in the spatiotemporal platform. Though LLI is not recovered in exact special relativistic terms, the alternative looks compatible with the relational aspects of general relativity (GR) with variable speed of light models as well a nonzero photon mass. This raises the hypothesis that Einstein equivalence principle, and consequently LLI, is a cornerstone of GR, but not necessarily a fundamental one of SR

Anisotropy in the speed of light plays a central role on the description of nature at the fundamental level while the link with metrology sets the importance of the observational status. Recent studies [1–7]  reveal controversial aspects of the fundamental constants and their variation, such as the fine structure constant, the electric charge, etc..., leading to possible variations of the fundamental couplings, to variable speed of light theories opposed to conventional pictures and faster than the speed of light signals. Since these models violate the Einstein equivalence principle at some level, search for possible Lorentz invariance violations become important, particularly the possibility of an anisotropic propagation velocity of light relative to a preferred reference frame . Following [8], for a laboratory moving with velocity *v* at an angle θ relative  to a preferred frame, the speed of light can be  written as:

$$\frac{c(\theta,v)}{c} = 1 + \left(\frac{1}{2} - \beta + \delta\right)\frac{v^2}{c^2}\sin^2\theta + (\beta - \alpha - 1)\frac{v^2}{c^2} \qquad (1)$$

where α = - ½ ,  β= ½ and δ = 0 in special relativity, reducing Eq. (1) to the first term.

Reported experimental tests [8,9] to determine Michelson-Morley as well as Kennedy-Thorndike coefficients,  $P_{MM} = (½ - β + δ)$  and $P_{KT} = (β - α - 1)$ sets the current limit to $| P_{MM} | ≤ 4.2 \times 10^{-9}$ and  $| P_{KT} | ≤ 6.9 \times 10^{-7}$, leading to Lorentz transformations confirmed to an overall uncertainty ~ $8 \times 10^{-7}$ , apart from determination of α from different tests [9] and the existence of additional terms in the standard model extension that describes Lorentz violations [4,8], such as first order in *v*/c. Here we investigate these additional terms and the uncertainty on the 'speed of light' measurements. The Lorentz factor, which is the root of Eq.(1) , can be written as,

$$\frac{1}{\sqrt{1 - \frac{v^2}{c^2}}} \approx 1 + \frac{1}{2}\frac{v^2}{c^2} + \frac{3}{8}\frac{v^4}{c^4} + \frac{5}{16}\frac{v^6}{c^6} + \ldots \tag{2}$$

Alternatively, from classical philosophical arguments [10], it was proposed [11] to distinguish the measured ($v_m$) velocity of any object from the effective velocity ($v$) the object is moving. This argument seems to result from a time gauge transform that includes the signal delay as:

$$\Delta t_m \rightarrow \Delta t + \Delta t_i \tag{3}$$

where $\Delta t_m$ is the measured time of an object evolution, $\Delta t$ it's effective or *proper time* and $\Delta t_i$ the signal time, or the time required for detecting the motion.

Employing the usual velocity definition in space-time ($v \equiv \Delta X / \Delta t$), a velocity gauge transform is obtained [11],

$$\frac{1}{v_m} = \frac{1}{v} + \frac{1}{v_S} \tag{4}$$

where $v_m$ is the measured velocity, $v$ the effective velocity and $v_S = 2c$ the signal velocity. This can be worked out giving a Lorentz–Einstein *Time Dilation* effect analogous expression,

$$\Delta t = \left(1 - \frac{v_m}{2c}\right)\Delta t_m \tag{5}$$

and an expression similar to Eq. (1),

$$v = \frac{v_m}{\left(1 - \frac{v_m}{2c}\right)} \tag{6}$$

provided $v$ in this last equation is interpreted as c ($\vartheta$), $\vartheta$ representing the variables of Eq. (1) and standing from possible additional terms not computed in the model described by Eq. (1). Adopting this terminology, the Lorentz analogous factor of Eq. (6) can be expanded leading to :

$$\frac{v}{v_m} \equiv \frac{c(\vartheta)}{c} \approx 1 + \frac{1}{2}\frac{v_m}{c} + \frac{1}{4}\frac{v_m^2}{c^2} + \frac{1}{8}\frac{v_m^3}{c^3} + \frac{1}{16}\frac{v_m^4}{c^4} + \ldots \tag{7}$$

This indicates that velocity measurements are always underestimated in spacetime, that is, c ( $\vartheta$ ) ≥ c for any finite signal velocity $v_S = 2c$, a result characteristic of the spacetime structure and independent of the existence of a preferred frame. Thus, odd $v_m/c$ terms ( Eq. (7)) could be present in optical experiments designed to test anisotropy, though they are not considered in the corresponding model of Eq. (1).

Recently, it was proposed [5] to call 'c' a "spacetime structure constant", once the speed 'c' involved in the local Lorentz transformations must be distinguished from the speed of light considered as a signal propagating in vacuum. In fact, Eq. (6) clearly distinguishes the measured velocity from the signal velocity, stating that the speed value (of any object's velocity) is a result of a balance ($v_m/v_S$) between the experimental data to the signal that carriers the data. Note that the classical condition corresponds to $v = v_m$ (not to $v_m = v_S$ ) and is equivalent to set $v_S \to \infty$ in Eq. (4). This should be also a relativistic condition ( or more generally a measurement condition ) where the signal is taken as a 'basis' or a *unit velocity vector*, in terms of which velocities are determined. Now, for experiments that somehow divide the incident beam and/or mixes signals, the magnitude of the 'reciprocal vectors' in Eq. (4) may become indistinguishable; if this drawback ( $v = v_S = v_m$ ) represents the overall current uncertainty ($\sim 10^{-7}$) for Lorentz violation tests we could argue that "null" results do not reflect the constancy of the "speed of light" but an experimental situation where the *spacetime structure constant*–c is mixed with a signal propagating in vacuum. In this sense, the gauge transformation–Eq. (4) belongs directly to the general theory of relativity, describing the relative motion of dynamical entities, in relation to one another, that is, a motion–to–signal ratio.

Another relevant point is the divergence of results: for instance, angle dependent effects, such as Michelson–Morley type tests, preserved about the same limit over two decades [9]. This picture is quite distinct when astrophysical sources are concerned. Gamma Ray Bursts provide a way to detect delays in light pulses traversing astronomical distances; reported data [12] registers $\Delta c / c$ ranging from $10^{-21}$ to $10^{-12}$ and photon masses $m_\gamma$ in the $10^{-44}$ to $10^{-35}$ g interval, depending on the event. Thus, though Einstein's postulate is preserved to such a level, the spread of data is quite different from the laboratory anisotropy tests and the question of whether there is a variation of the speed of light in vacuum is open; moreover, in the absence of a conclusive c(ω) model [13,14], including cavity tests, the phenomenological approach relates the relative spread $\Delta c / c$ to the  (astronomical) distance D, a 'c' dependent figure. This dependence should not affect the relative speed accuracy but the absolute c values may depend on the motion–to–signal ratio above discussed.

Finally, the photon mass plays a central role either as an attributable source of the c (ω) variation but mainly as a fundamental physical entity; however, SR adoption of $m_\gamma = 0$  rests on the energy relation and can be considered a misinterpretation of  the kinematical relativistic equations [15]. Here we note that while the Lorentz transformation factor–  Eq.(2) is closely connected to the *photon rest mass* definition, the gauge transformation alternative term acts on times–Eq.(4) and on velocities–Eq.(6), but not on masses; this means the same treatment of SR as far as energy–momentum relations are concerned, including similar figures, but the possibility of  nonzero photon masses and a variation of the speed of light trough a dependence on the signal carrying the information.

## Acknowledgments

Useful discussions with M Barreira are gratefully ackowledged